\documentclass[runningheads]{svmult}

\usepackage{makeidx}   
\usepackage{graphicx}  
\usepackage{subeqnar}  
\usepackage{multicol}  
\usepackage{physprbb}  
\makeindex             

\begin{document}
\title*{On the Fast Spectral Variability of GRBs}
\toctitle{On the Fast Spectral Variability of GRBs}
%
%
\titlerunning{On the Fast Spectral Variability of GRBs}
%
\author{E.P.Mazets\inst{1}
\and R.L.Aptekar\inst{1}
\and P.S.Butterworth\inst{2}
\and T.L.Cline\inst{2}
\and D.D.Frederiks\inst{1}
\and S.V.Golenetskii\inst{1}
\and V.N.Il'inskii\inst{1}
\and V.D.Pal'shin\inst{1}}
\authorrunning{E.P.Mazets et al.}
%
%
\institute{Ioffe Physico-Technical Institute, St.Petersburg, 194021, Russia
\and Goddard Space Flight Center, Greenbelt, MD 20771, USA}

\maketitle              
\vspace*{-0.3cm}
\begin{abstract}
Fast spectral variability of gamma-ray burst emission is considered 
for a number of events seen by the Konus-Wind experiment. 
The variability manifests itself as a strong
correlation between instantaneous energy flux $F$ and peak energy $E_p$.
In the ($F,E_p$) plane, the correlation produces distinct tracks 
in the form of branches and loops representing the different parts of a burst time history.
Despite the variety of features seen in different events, the main
characteristics of the spectral evolution produce a quite consistent pattern.
\end{abstract}
\vspace*{-0.3cm}
The temporal evolution of GRB energy spectra has been clearly seen in first experiments
with sufficient spectral and time resolution.
Some characteristics of spectral evolution were pointed out in the Konus experiment
aboard Venera 11 and Venera 12~\cite{Mazets81}.
Observations on Venera 13 and 14 revealed a strong correlation between
instantaneous intensity and hardness of a radiation~\cite{Gol83}.
Then, many authors considered different aspects of spectral evolution of
gamma-ray bursts~\cite{Norris86,Kargatis94}. 
The BATSE experiment aboard the CGRO offers a wealth of spectroscopic
information permitting a rich study of the variability and related 
correlations in GRBs~\cite{Ford95,Borgonovo01}.
The Konus-Wind gamma-ray burst experiment flown onboard the GGS WIND spacecraft~\cite{Aptekar95}
has sampled about 900 bursts since lunch in 1994. A sufficient 
part of this sample consists of events which are strong enough that
we can further extend studies of spectral evolution.

The Konus instrument records a time history and some prehistory of a burst in three
energy windows: G1=10--50~keV, G2=50--200~keV, G3=200--750~keV (nominal values)
with a time resolution of 2--256~ms, and also up to 64 multichannel energy spectra
in the 10~keV--10~MeV range with accumulation times of 64~ms--8~s.
Energy loss spectra obtained are deconvolved to incident photon spectra
using the detector response function for known angle of incidence.
For fitting the spectra, we used the empirical model by Band with spectral parameters
$\alpha,\beta,E_0$~\cite{Band93}.
Thus, for each energy spectrum we obtain values of an energy flux $F$
and a peak energy $E_p=E_0(2-\alpha)$.
These data allows the consideration of a spectral evolution with
moderate time resolution.
Finer time resolution can be obtained by considering
time history measurements.
The three time profiles G1, G2, G3 give two independent
sets of hardness ratio measurements, for example G2/G1, and G3/G2.
Each pair of these values depends on spectral parameters
corresponding to a short time interval. We can calculate
expected hardness ratios using the response function matrix and the incident photon
spectrum with parameters $\alpha,\beta,E_0$.
In each time bin, the best fit parameters are determined by minimizing 
the quadratic sum of the two weighted differences between the calculated and 
observed hardness ratio values.
This procedure is illustrated with spectral evolution data for GRB~950822 in Fig.~1.
This figure displays a lot information.
The mutual dependence of the instantaneous values of the energy flux $F$
and the hardness of a spectrum $E_p$ forms a distinct track in the ($F,E_p$).
The track consists of three main branches. The first branch, 0.2 to 7.5~s,
represents a weak but very hard emission in initial stage of the burst.
The second branch, 7.5 to 13~s, corresponds to the rising of the main pulse
of the burst. Finally the long branch for burst decay, 13 to 30~s,
ends with very soft emission.
Each branch can be approximated by a power law relation 
$F\propto E_p^\gamma$ where indices $\gamma$ are correspondingly
$\sim0.7,\sim6,$ and $\sim2.5$.
Such spectral behaviour is typical of a majority
of gamma-ray bursts, especially bursts with well separated pulses.
Similar branches and loops are seen in many events. 
A number of further examples are presented in Fig.~2.

Only count rate statistics constrain
the time resolution with which correlation can be observed.
This is seen in Fig.~2j which presents the correlation tracks of very short ($\sim$100~ms)
and hard ($E_\gamma >2$~MeV) GRB~970704 obtained with a time resolution of 2~ms.

Finally, it should be noted, that the SGR~1627-41 also exhibits a strong spectral
variability~\cite{Aptekar}.
Fig.~2l demonstrates that the correlation tracks of SGR if shifted
along the hardness axis $E_p$ by one order of magnitude look like
the tracks for GRBs.

A more complete description of the Konus-Wind data on the spectral
evolution in GRBs will be published elsewhere.

\begin{figure}[!t]
\vspace*{-0.3cm}
\parbox[]{0.7\textwidth}%
{\includegraphics*[width=0.69\textwidth,bb=130 377 475 790]{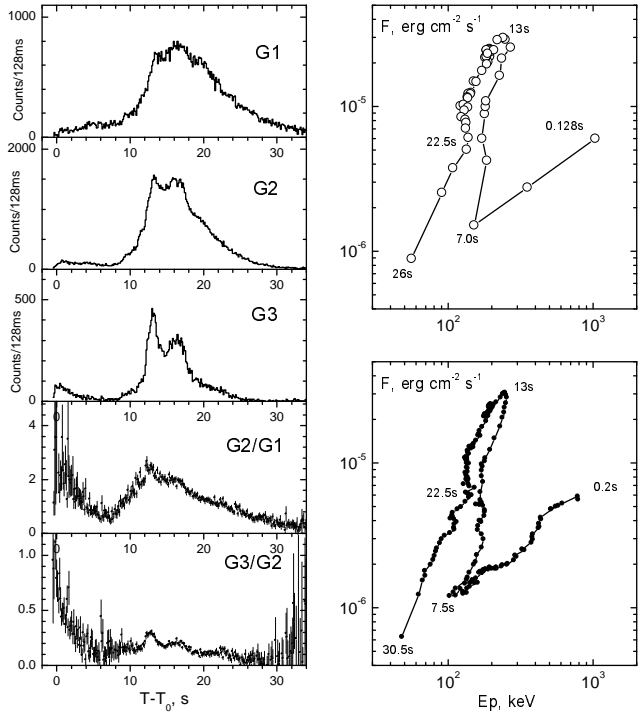}}
\hfill
\parbox[]{0.29\textwidth}%
{\caption{The spectral evolution in GRB~950822. Left panel: time history
and hardness ratio time profiles. Right panel (top): the correlation
between intensity and hardness. Data are obtained from energy spectra measurement.
Right panel (bottom): Intensity-hardness correlation.
Data are obtained using the hardness ratio profiles.
Numbers placed along tracks correspond to the times $T-T_0$
on the time history profile.}}\\
\vspace*{-0.6cm}
\end{figure}

\clearpage

\begin{figure}[!h]
\vspace*{-0.2cm}
\centering\includegraphics*[height=13.2cm,bb=103 268 433 735]{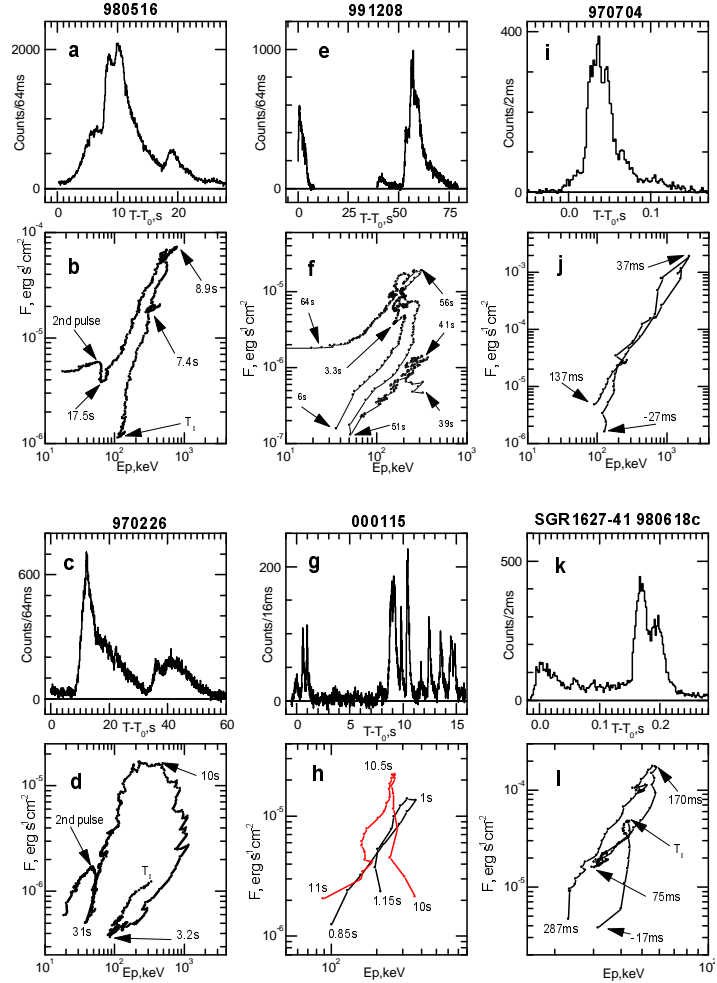}\\
\vspace*{-0.2cm}
\caption{Time histories and correlation tracks of five gamma-ray bursts
and one recurrent burst from SGR~1627-41. Panel \textbf{h}: Only two tracks
for two pulses are shown for clarity.}
\vspace*{-0.6cm}
\end{figure}
%
This work was supported by Russian Aviation and Space Agency Contract, 
RFBR grant N 99-02-17031 and CRDF grant RP1-2260.

%


\vspace*{-0.4cm}


\begin{thebibliography}{8.}
%
\vspace*{-0.2cm}
\bibitem{Mazets81} Mazets, E.P., et al.: Ap\&SS, \textbf{75}, 47 (1981)
\bibitem{Gol83} Golenetskii, S.V.,et al.: Nature, \textbf{306}, 451 (1983)
\bibitem{Norris86} Norris, J.P., et al.: ApJ, \textbf{301}, 213 (1986)
\bibitem{Kargatis94} Kargatis, V.E., et al.: ApJ, \textbf{422}, 260 (1994)
\bibitem{Ford95} Ford, L.A., et al.: ApJ, \textbf{439}, 307 (1995)
\bibitem{Borgonovo01} Borgonovo, L., \& Ryde, F.: ApJ (2001) in press, astro-ph/0009164
\bibitem{Aptekar95} Aptekar, R.L., et al.: Space Science Rev., \textbf{71}, 265 (1995)
\bibitem{Band93} Band, D., et al.: ApJ, \textbf{413}, 281 (1993)
\bibitem{Aptekar} Aptekar, R.L., et al.: These Proceedings.
%
\end{thebibliography}
\end{document}